\documentstyle[twocolumn,pra,aps,psfig]{revtex}

\begin{document}
\twocolumn[\hsize\textwidth\columnwidth\hsize\csname@twocolumnfalse\endcsname
\draft
\title{No enhancement of the localization length \\ for two
  interacting particles in a random potential}

\author{Rudolf A. R\"{o}mer and Michael Schreiber}

\address{ Institut f\"{u}r Physik, Technische Universit\"{a}t
  Chemnitz, D-09107 Chemnitz,
  Germany }

\date{Version: October 24, 1996; printed \today} \maketitle

\begin{abstract}
  We study two interacting particles in a random potential chain by
  means of the transfer matrix method.  The dependence of the
  two-particle localization length $\lambda_2$ on disorder and
  interaction strength is investigated.  Our results demonstrate that
  the recently proposed enhancement of $\lambda_2$ as compared to the
  results for single particles is entirely due to the finite size of
  the systems considered.  This is shown for a Hubbard-like onsite
  interaction and also a long-range interaction.
\end{abstract}

]

%
%


In one dimension (1D), bosonization of the repulsive Hubbard model in
the gapless phase shows that (onsite) disorder is a relevant
perturbation and the ground state corresponds to a strongly localized
phase \cite{schulz1}.  Thus it came as a surprise, when Shepelyansky
\cite{shep1} recently argued that the Hubbard interaction between two
particles in a random potential would reduce the localization in
comparison with independent particles.  In particular, he obtained an
enhancement of the two-particle localization length $\lambda_2$
independent of the statistics of the particles and of the sign of the
interaction such that
\begin{equation}
  \lambda_2 \approx U^2 \frac{\lambda_1^2}{32}
\label{eq-shep}
\end{equation}
in the band center. Here $\lambda_1$ is the single particle
localization length in 1D and $U$ the Hubbard interaction in units of
the nearest-neighbor hopping strength.  Shepelyansky obtained this
result by studying the matrix representation ${\bf U}$ of the Hubbard
interaction in the disorder-diagonal basis of localized single
particle eigenstates, i.e., ${\bf U} = U Q_{nm,kl}= U \sum_{i}
\phi^\dagger_n(i) \phi^\dagger_m(i) \phi_k(i) \phi_l(i)$, with
$\phi_n(i)$ the amplitude at site $i$ of the single particle
eigenstate with energy $E_n$.  Here, he assumed that $\phi_n(i)$
behaves as $\phi_n(i) \sim r_n/\sqrt{\lambda_1}$, where $r_n$ is a
random number of order unity, and he neglected correlations among the
$r_n$ at different sites resulting in a Gaussian distribution for the
matrix elements of $Q$.

Support for this result was given shortly afterwards by Imry
\cite{imry1} with the help of a Thouless-type scaling argument.  Frahm
{\em et al}.\ \cite{frahm1} used the transfer matrix method (TMM) to
study the two-particle Hubbard-Anderson problem without any
approximations and have found numerically that $\lambda_2 \sim
\lambda_1^{1.65}$.  They also measured the distribution of the matrix
elements of $Q$ and found a strongly non-Gaussian behavior.  They
attribute the deviation of the exponent from Eq.~(\ref{eq-shep}) and
the different distribution of matrix elements to the correlations
among the $r_n$ neglected by Shepelyansky.
Subsequently, an approximate calculation of $\lambda_2$ by Oppen {\em
  et al}.\ \cite{oppen1} with Green function methods lead to the
hypothesis $\lambda_2 = \lambda_1 + C \frac{|U|}{\pi} \lambda_1^2$,
with $C\approx 0.34$ for bosons and $0.36$ for fermions.  Oppen {\em
  et al}.\ also identified a scaling parameter $U \lambda_1$.
Recently, Weinmann {\em et al}.\ \cite{wein2} argued that for a
Gaussian matrix ensemble there exists a crossover from the $|U|$
behavior to the $U^2$ behavior for increasing $U$, but with a
different exponent, such that $\lambda_2 = \lambda_1 + A
\frac{|U|}{4+W} \lambda_1^{3/2}$ for small $U$ with $W$ parametrizing
the disorder.

\begin{figure}
  \centerline{\psfig{figure=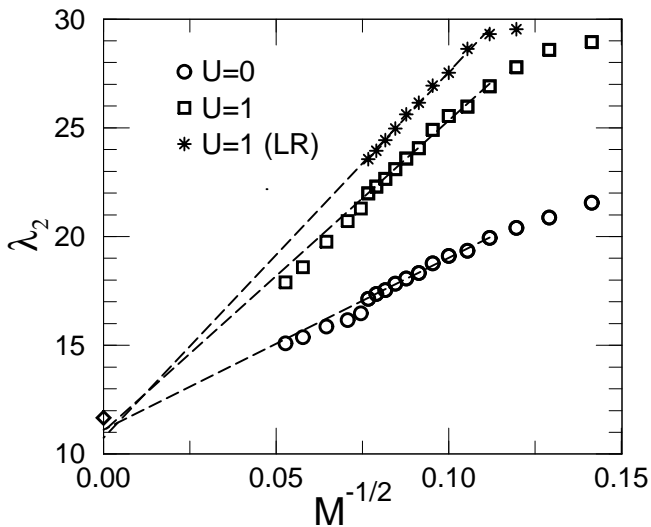,width=3.6in,height=3.0in}}
  \caption{
    Two-particle localization length $\lambda_2$ as a function of the
    system size at $W=3$ for Hubbard ($\Box$) and long-range
    interacting systems ($*$).  Dashed lines are extrapolations from
    the data for $M\in [80,170]$.  The $5$ leftmost data points for
    $U=0$ and Hubbard $U=1$ corresponding to $M=180, 200, 240, 300,
    360$ have been computed with only $5\%$ accuracy and have not been
    considereded for the extrapolations.  The diamond ($\diamond$)
    indicates the value of $\lambda_1$.  }
\label{fig-l2_m}
\end{figure}

We feel that in this ongoing controversy about the interaction and
disorder dependence, the system size has not been appropriately taken
into consideration. We doubt the existence of the enhancement for an
infinite system size, because there is only a vanishing probability
that two localized particles are sufficiently close on an infinite
chain to ``see'' each other. To substantiate this argument, we present
in this Letter extensive computations of $\lambda_2$ in dependence of
the system size $M$. We first review the TMM approach to the two
interacting particles (TIP) problem and suggest an improved TMM based
on the self-averaging properties of the Lyapunov exponents.  With this
method we reproduce the $\lambda_2$ estimates of Ref.~\cite{frahm1}
for $U=1$, system size $M=100$, and all disorders.  We find (cp.\ 
Fig.~\ref{fig-l2_m}), however, that the enhancement
$\lambda_2/\lambda_1$ decreases with increasing $M$.  We also study
the behavior of $\lambda_2$ for $U=0$ and show in Fig.~\ref{fig-l2_m}
that it is not equal to $\lambda_1$ for finite $M$.  But we recover
the $\lambda_1$ value \cite{numtmm} in the limit $M\rightarrow\infty$.
However, the enhancement $\lambda_2/\lambda_1$ also vanishes
completely in this limit as demonstrated in Fig.~\ref{fig-l2_m}.

%
%


The Schr\"{o}dinger equation for the TIP problem with Hubbard
interaction is written in a suggestive form as
\begin{eqnarray}
  \psi_{n+1,m} &= & \left[ E - (\epsilon_n+\epsilon_m) - U
    \delta_{n,m} \right] \psi_{n,m} \nonumber \\ & & \mbox{ } -
  \psi_{n,m+1} - \psi_{n,m-1} - \psi_{n-1,m},
\label{eq-tip}
\end{eqnarray}
where $n, m= 1, \ldots, M$ are the two site indices of the particles,
$E$ is the total energy of both particles, and $\epsilon_i$ is the
random potential at site $i$.
In the following, we use a box distribution $[-W/2,W/2]$ for the
$\epsilon_i$.  The single particle localization length in 1D for such
a distribution is known \cite{thouless} from second order perturbation
theory in $W$ to vary with disorder and energy as
$
\lambda_1 \approx 24( 4 - E^2 )/W^2 $.
The numerical results for $E=0$ based on the TMM \cite{numtmm} lead to
a slightly different prefactor
\begin{equation}
  \lambda_1 \approx 105/W^2.
\label{eq-l1}
\end{equation}


If one interprets $(n,m)$ as Cartesian coordinates on a finite lattice
with $M\times M$ sites, the problem becomes identical to a
non-interacting Anderson model in 2D with disorder potential symmetric
with respect to the diagonal $n=m$ and with {\em hard wall} boundary
conditions \cite{frahm1,mackinnon1}.  One can rewrite
Eq.~(\ref{eq-tip}) in the TMM form as
\begin{equation}
  \left( \begin{array}{l} \psi_{n+1} \\ \psi_{n} \end{array} \right) =
  T_n \left( \begin{array}{l} \psi_{n} \\ \psi_{n-1} \end{array}
  \right)
\label{eq-loctmm}
\end{equation}
with the symplectic transfer matrix
\begin{equation}
  T_n= \left(
 \begin{array}{cc}
   E \openone - \chi_n - H_{\perp} \quad & - \openone \\ \openone &
   {\bf 0}
 \end{array}
\right),
\end{equation}
where $\psi_n= (\psi_{n,1}, \ldots, \psi_{n,m}, \ldots, \psi_{n,M})$
is the wave vector of slice $n$, $H_{\perp}$ is the single-particle
hopping term for the transverse ($m$) electron and $(\chi_n)_{i,m}=
(\epsilon_n + \epsilon_m + U \delta_{n,m}) \delta_{i,m}$ codes the
random potentials and the Hubbard interaction \cite{frahm1}.  Note
that in this approach the symmetry of the wave function remains
unspecified and we cannot distinguish between boson and fermion
statistics.

The evolution of the state is determined by the matrix product
$\tau_N= \prod_{n=1}^{N} T_n$ and we have
\begin{equation}
  \left( \begin{array}{l} \psi_{N+1} \\ \psi_{N} \end{array} \right) =
  \tau_N \left( \begin{array}{l} \psi_{1} \\ \psi_{0} \end{array}
  \right).
\label{eq-globtmm}
\end{equation}
Usually, the method is performed with a complete and orthonormal set
of initial vectors $(\psi_1,\psi_0)^T$.
The eigenvalues $\exp[2 \gamma_i(N)]$ of
$(\tau^{\dagger}_N\tau_N)^{1/N}$ exist for $N\rightarrow\infty$ due to
Oseledec's theorem \cite{oseledec}.
The smallest Lyapunov exponent $\gamma_{\rm min}$ determines the
slowest possible decay of the wave function and thus the largest
localization length $\lambda_{\rm max}= 1/\gamma_{\rm min}$.  We now
{\em define} the two-particle localization length $\lambda_2$ as
$\lambda_{\rm max}$ of the transfer matrix problem (\ref{eq-globtmm}).

In the TMM studies of the Anderson Hamiltonian, one usually multiplies
the transfer matrices in the $n$ direction until convergence is
achieved, e.g., for a 2D sample with strip width $M=100$ and $W=2$,
this typically requires $N=5\times 10^6$ multiplications for an
accuracy of $1\%$. Thus one effectively studies a quasi-1D system of
size $M \times N$ with $M\ll N$.  However, in the present problem,
both directions are restricted to $n,m \leq M$.  Iterating
Eq.~(\ref{eq-loctmm}) only $M$ times will not give convergence.
Frahm {\em et al}.\ \cite{frahm1} have solved this problem in their
TMM study by exploiting the hermiticity of the product matrix $Q_M=
\tau^\dagger_M \tau_M$: Continuing the iteration (\ref{eq-globtmm})
with $\tau^\dagger_M$, then with $\tau_M$, and so on, until
convergence is achieved, yields the eigenvalues $\exp[2 \gamma_i(N)]$
of $( \prod^{N/M} Q_M)^{1/N}$.  This means that they repeat the {\em
  same} $M\times M$ sample $N/M$ times.  For $M=100$ and $W=2$ we
found that $N/M\approx 5000$ is necessary to give an accuracy of
$5\%$.  Since this has to be done for every disorder configuration,
the computational effort is quite large.
However, in the usual quasi-1D TMM for single particles, one may also
halt the matrix multiplication every $L\ll N$ steps, compute an
estimate for $\gamma(L)$ and finally average over all $N/L$ such
values \cite{kramer1}.  Due to the self-averaging property of the
Lyapunov exponents \cite{selftmm} the averaged localization length
equals the converged localization length $\lambda(N)$ within the
accuracy of the calculation.  This suggests the following method: We
iterate $\prod_{r=1}^{N/M} \tau_{r,M}$, where $r$ enumerates {\em
  different} samples of size $M \times M$.  Next we compute a
localization length $\lambda(N)$ which is now already averaged over
$N/M$ different samples.  This significantly reduces the computational
effort and allows us to study much larger systems and achieve a
significantly better accuracy than previously.

%
%



\begin{figure}
  \centerline{\psfig{figure=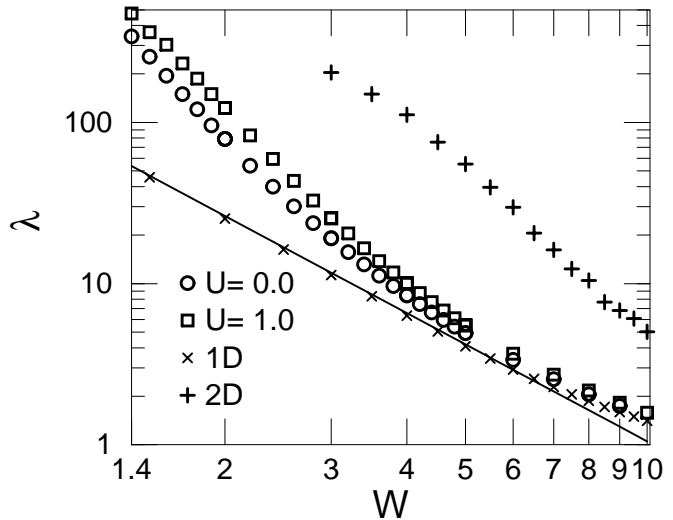,width=3.6in,height=3.0in}}
  \caption{
    Two-particle localization length $\lambda_2$ (accuracy at least
    $1\%$) at energy $E=0$ for system size $M=100$ and interaction
    strength $U=0$ ($\Large \circ$) and $1$ ($\Box$).  The solid line
    is given by Eq.~(\ref{eq-l1}).  Single-particle TMM data with
    $1\%$ accuracy for 1D chains ($\times$) and 2D strips of width
    $M=100$ ($+$) are also shown.  }
\label{fig-l2_w}
\end{figure}

In Fig.~\ref{fig-l2_w} we show the results obtained for the
localization length of the TIP problem with Hubbard interaction by the
present TMM.
The data is calculated by configurationally averaging at least $10
000$ samples for each data point.  This corresponds to an accuracy of
$1\%$ for the smallest $W$ considered and better for larger $W$.  We
note that the data for $U=1$ agree well with the results obtained in
Ref.~\cite{frahm1} for $1.4\leq W \leq 4$ \cite{differentw} and we
find $\lambda_2\sim W^{-3.8}$.  Furthermore, for $W\in [1.4,5]$, the
data can also be fitted reasonably well by
\begin{equation}
  \lambda_2 = \lambda_1 + A \lambda_1^\alpha / (B + W),
\label{eq-l2l1}
\end{equation}
with $\alpha=2$ with fit parameters $A=0.37$ and $B=0.7$.  A fit with
$\alpha=3/2$ according to Ref.~\cite{wein2} is considerably worse and
without the denominator in Eq.~(\ref{eq-l2l1}), as suggested in
Ref.~\cite{oppen1}, the data cannot be fitted at all.
We emphasize that the TIP data differ significantly from the 2D TMM
data with uncorrelated 2D random potential $\epsilon_{n,m}$.

In Fig.~\ref{fig-l2_w} we have also plotted the behavior for $U=0$,
where the system reduces to two non-interacting particles in 1D.
Before analyzing the data, let us first state our expectations: For
$U=0$, the two particles will localize independently at two arbitrary
sites, say $n_0$, $m_0$, with localization length $\lambda_1$.  The
wave function can then be written as a product of two exponentially
decaying single particle wave functions, i.e.  $\psi^{(0)}_{n,m}\sim
\exp [-|n-n_0|/\lambda_1(E_0)] \exp [-|m-m_0|/\lambda_1(-E_0)]$.
Here, the eigenenergies are chosen such that $E=0$.  Two points are
worth mentioning: (i) Let $\gamma_1(E_\alpha)$ denote the inverse
decay length measured by TMM for a single particle with energy
$E_\alpha$ and $\gamma_2(0)$ the inverse decay length for two
particles with $E= 0$.  For $U=0$, the TMM for two particles will
measure an inverse decay length $\gamma_2 (0) \approx
\sum_{\alpha}^{\cal N} \gamma_1(E_\alpha) / {\cal N}$, averaged over
${\cal N}$ pairs of states with energies $E_\alpha$ and $ - E_\alpha$.
Since $\gamma_1(E\neq 0) > \gamma_1(0)$ as mentioned above
Eq.~(\ref{eq-l1}), we have $\gamma_2 (0) > \gamma_1(0)$ and
consequently $\lambda_2(0) < \lambda_1(0)$.  (ii) The non-interacting
two-particle wave function $\psi^{(0)}_{n,m}$ is not isotropic in the
2D plane $(n,m)$. Since the TMM will not necessarily measure the decay
directly in the $n$ direction, we expect that $\gamma_2$ will also
contain information about the decay in other directions.  These decay
lengths of $\psi^{(0)}_{n,m}$ are shorter and thus we again expect
$\lambda_2(0) \leq \lambda_1(0)$.

Figure \ref{fig-l2_w} shows that contrary to our expectations the TMM
does not under- but overestimate $\lambda_2$ as compared to
$\lambda_1$ such that the enhancement $\lambda_2(U=1)/\lambda_2(U=0)$
is much smaller than the previously reported
$\lambda_2(U=1)/\lambda_1$.  Also included in Fig.~\ref{fig-l2_w} is a
1D TMM result and we see that for $W\geq 6$ --- where the perturbative
result (\ref{eq-l1}) is no longer accurate --- both the $U=0$ and the
$U=1$ data agree quite well with $\lambda_1$.  For $W\in [1.4,4]$ a
power law fit gives $\lambda_2(U=0)\sim W^{-3.6}$.  A fit with
$\alpha=2$ as in Eq.~(\ref{eq-l2l1}) can also describe the data for
small $W$ with $A=0.19$ and $B=0.5$.

We emphasize that the deviation from Eq.~(\ref{eq-l1}) of our data for
vanishing $U$ does not result from an effective coupling which was
suspected \cite{amg1} to be introduced into the problem because of
numerical instabilities of the TMM.  Our tests did not show such a
coupling.  We also note that the two-particle TMM data is quite
different from data of a 2D single-particle calculation for $M=100$ as
shown in Fig.\ \ref{fig-l2_w}.



\begin{figure}
  \centerline{\psfig{figure=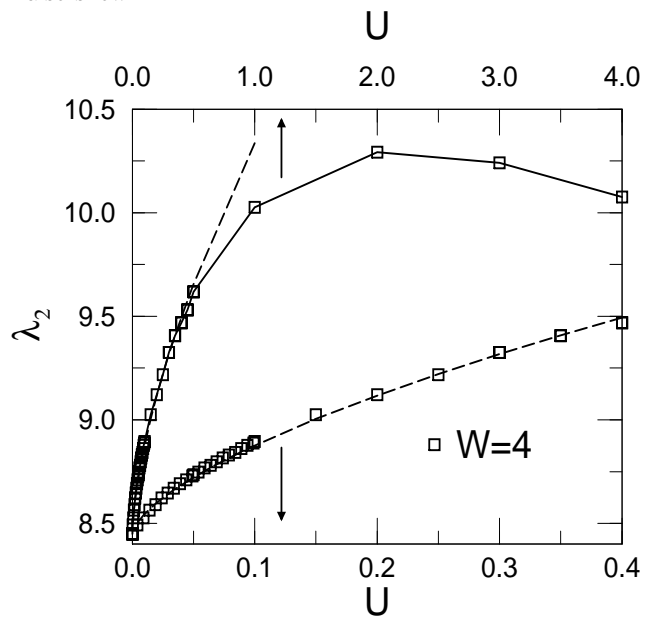,width=3.4in,height=3.0in}}
  \caption{
    $U$ dependence of $\lambda_2$ for $W=4$. The dashed line
    represents the power law fit $\lambda_2(U) - \lambda_2(0) \approx
    U^{0.63}$.  The lower curve shows the same data on an enlarged $U$
    scale.  Arrows indicate the corresponding axes.  }
\label{fig-l2_u}
\end{figure}

Studying the $U$ dependence of $\lambda_2$, we note that
$\lambda_2(U)=\lambda_2(-U)$ at $E=0$ due to particle-hole symmetry.
Consequently, we only show data for $U\geq 0$ in Fig.~\ref{fig-l2_u}.
We observe neither the proposed $U^2$ nor the $|U|$ behavior.  Rather
the data obey $\lambda_2(U) - \lambda_2(0) \approx U^{0.63}$ for small
$U$.  Also, $\lambda_2(0)\approx 8.45$ does not agree with
$\lambda_1\approx 105/16$, as discussed above, and we cannot identify
a scaling parameter as suggested in Ref.~\cite{oppen1}.


{}From Fig.~\ref{fig-l2_w} we see that for $W > 6$, the data for $U=0$
and $U=1$ coincide with the single-particle TMM data already quite
well.  The enhancement $\lambda_2/\lambda_1$ is large only for small
disorder ($W\leq 4$).  However, in this region, the computed values of
$\lambda_2$ become comparable to the size of the system such that an
increasing part of the wave function is reflected at the hard wall
boundaries. This leads to an overestimation of $\lambda_2$.
In order to distinguish this artificial from the proposed enhancement
due to interaction alone, we have therefore systematically studied the
finite-size behavior of $\lambda_2$ at various disorder values and
interaction strengths.  As an example, we consider the case $W=3$.
Here the localization lengths for $\lambda_2(U=0)\approx 19.1$ and
$\lambda_2(U=1)\approx 25.5$ may just be small enough for the wave
function to fit into the $100\times 100$ box without too strong
boundary effects.  Furthermore, the enhancement is still sufficiently
large to be numerically detectable.
In Fig.~\ref{fig-l2_m}, we have shown $\lambda_2$ as a function of
$M^{-1/2}$.  Our reason for plotting the data as a function of
$M^{-1/2}$ is pragmatic: a plot versus $M^{-1}$ or $M^{-1/4}$ shows a
downward or upward curvature, respectively, for the largest $M$.  The
curvature is smallest for $M^{-1/2}$ and the extrapolation towards
infinite $M$ is most reliable.  We see in Fig.~\ref{fig-l2_m} that (i)
the data for $U=0$ approach the single-particle result $\lambda_1
\approx 105/9$ for large $M$, and, more significantly, (ii) the data
for $U=1$ also approach this non-interacting 1D result.  Therefore the
overestimation of $\lambda_2(U=0)$ vanishes for large $M$, supporting
the validity of the TMM approach.  However, in this limit the
enhancement $\lambda_2(U=1)/\lambda_1$ vanishes completely, too.
We note that the observed decrease in the localization length $\lambda_2$
is quite different from a 2D TMM, where the localization length is
known to increase with increasing system size \cite{mackinnon1}.  We
attribute this difference to the symmetry of the TIP disorder
potential which is equivalent to an effective long-range correlation
which furthermore increases with increasing $M$.

%
%


%
%


We have also studied the behavior of the two-particle localization
length for a system with long-range interaction, since Shepelyansky
argues in Ref.~\cite{shep3} that long-range interaction should also
lead to the enhancement of $\lambda_2$.  The Schr\"{o}dinger equation
is given as
\begin{eqnarray}
  \psi_{n+1,m} &= & \left[ E - (\epsilon_n+\epsilon_m) - U/(|n-m|+1)
  \right] \psi_{n,m} \nonumber \\ & & \mbox{ } - \psi_{n,m+1} -
  \psi_{n,m-1} - \psi_{n-1,m} ,
\label{eq-tip-lr}
\end{eqnarray}
such that the on-site interaction is equal to the Hubbard interaction.
Clearly such a system should exhibit even stronger boundary effects
and we thus expect an even stronger enhancement of $\lambda_2$ for
finite systems.
In Fig.~\ref{fig-l2_m}, we have also shown $\lambda_2$ for the
long-range interaction with $U=1$ at $E=0$ and the same accuracy of
$1\%$ as for the Hubbard interacting system. As expected, we observe
that the enhancement for all $M$ is stronger than for the Hubbard
interacting model. But this enhancement again vanishes for
$M\rightarrow\infty$.

%
%


In summary, we have studied the interaction-induced enhancement of the
localization length for two particles in a 1D random potential by
means of an improved TMM.  For $U=1$ and $M=100$, the results agree
well with previously published data. However, the $U=0$ data is quite
surprising since only in the limit $M\rightarrow\infty$ do we recover
the expected non-interacting 1D result $\lambda_1$.  However, in this
limit the enhancement for finite $U$ also vanishes.
Therefore we must conclude that the transfer matrix method applied to
the 1D Anderson model for two interacting particles measures an
enhancement of the localization length which is entirely due to the
finiteness of the systems considered. This enhancement might be
relevant for applications in mesoscopic systems.

\acknowledgments We thank Frank Milde for programming help and
discussions.  This work has been supported by the Deut\-sche
For\-schungs\-ge\-mein\-schaft as part of
Son\-der\-for\-schungs\-be\-reich 393.

\begin{references}
\bibitem{differentw} Note that in Ref.~\protect\cite{frahm1} the
  disorder distribution $[-W,W]$ is used.
\end{references}
\end{document}